\documentclass[twocolumn,amsmath,amssymb]{article}

\makeatletter
\def\@normalize{@setsize\normalsize{12pt}\xpt\@xpt
\abovedisplayskip 10pt pluse2pt minus5pt\belowdisplayskip
\abovedisplayskip \abovedisplayshortskip \z@ plus3pt\belowdisplayshortskip
6pt plus3pt minus3pt\let\@listi\@listI}
\def\subsize{\@setsize\subsize{12pt}\xipt\@xipt}
\def\section{\@startsection{section}{1}{\z@}{24pt plus 2 pt minus 2 pt} {12pt plus 2pt
minus 2pt}{\large\bf}}
\def\subsection{\@startsection{subsection}{2}{\z@}{12pt plus 2 pt minus 2 pt}{12pt plus
2pt minus 2pt}{\subsize\bf}}
\makeatother

\usepackage{graphicx} 
\usepackage{dcolumn} 
\usepackage{bm} 

\begin{document}
\date{}
\title{\large\bf Effects of feedback and feedforward loops on dynamics\\ of transcriptional regulatory model networks}
\author{Chikoo Oosawa$^{1,}\footnote{Corresponding author: chikoo@bio.kyutech.ac.jp}$ , Kazuhiro Takemoto$^{2}$, Michael A. Savageau$^{3}$\\\\$^{1}$Department of Bioscience and Bioinformatics, Kyushu Institute of Technology \\ Iizuka, Fukuoka 820-8502, Japan \\ $^{2}$Bioinformatics Center, Institute for Chemical Research, Kyoto University \\ Gokasho, Uji, Kyoto 611-0011, Japan \\$^{3}$Department of Biomedical Engineering, College of Engineering \\University of California, Davis, 95616, U.S.A.}

\maketitle
\thispagestyle{empty}
\subsection*{\centering Abstract}
\vspace*{-3mm}
We demonstrate the advantages of feedforward loops using a Boolean network, which is one of the discrete dynamical models for transcriptional regulatory networks. After comparing the dynamical behaviors of network embedded feedback and feedforward loops, we found that feedforward loops can provide higher temporal order (coherence) with lower entropy (randomness) in a temporal program of gene expression. In addition, complexity of the state space that increases with longer length of attractors and greater number of attractors is also reduced for networks with more feedforward loops. Feedback loops show opposite effects on dynamics of the networks. These results suggest that feedforward loops are one of the favorable local structures in biomolecular and neuronal networks.\\
\\Keywords : Boolean networks; feedback loop; feedforward loop; mutual information; entropy; transcriptional regulatory networks
\section{Introduction}
Recent studies of natural complex networks \cite{BOOK} including transcriptional regulatory networks in cells have revealed at least three stastistical properties: long-tailed global connectivity distributions having a small number of highly connected nodes; small-world properties of short path lengths between any two nodes; and highly clustered connections among adjacent nodes \cite{NETBIO,Bridge,WATTS,JEONG00,KASHTAN,Milo,Ishihara,Alon}. The last local structures called motifs consist of a few nodes and edges among the nodes which are found to be statistically significant, and can be regarded as functional modules \cite{KASHTAN,Milo,Ishihara}. Since feedback and feedforward loops are motif structures as well as basic and ubiquitous circuits in man-made systems, one can expect that transcriptional regulatory networks also have both feedback and feedforward loops; however, only the feedforward loops prevail \cite{KASHTAN,Milo,Ishihara}. Other biological networks such as signal transduction and neuronal networks also have similar tendencies, which suggests that feedforward loops are favored in complex biological networks. In general, although massive available network data demonstrate the statistical significance, it is unclear why feedforward loops are advantageous over feedback loops in dynamical systems.\\
\section{Model and method}
\subsection{Boolean network}
The dynamics of the Boolean networks \cite{PHD02,SAK93} is determined by the equation
\begin{equation}
X_i(t+1)=B_i\left[\bm{X}(t)\right]\quad(i=1,2,...,N),
\label{eq:bn}
\end{equation}
where $\bm{X}(t)$ is a binary state, either 0 or 1, of node {\it i} at time $t$, $B_i(\cdot)$ are Boolean functions [See Table \ref{tab:bool}] used to simultaneously update the state of node {\it i}, and $\bm{X}(t)$ is a binary vector that gives the states of the $N$ nodes in the network. After assigning the initial states $\bm{X}(0)$ to the nodes, the successive states of the nodes are updated by input states and their Boolean functions. The dynamical behavior of these networks is represented by a time series of binary states. The time course follows a transient phase from an initial state until a periodic pattern, called an attractor, is established [See Fig. \ref{fig:state-space}].
\begin{table}[hbt]
\caption{4 of 16 Boolean functions with indegree $K_{in}$ = 2. In this paper, we used Boolean functions shown below, because of the feasibility of computation and the biological meaning of the functions \cite{LRAEY02,Alon03}.}
\begin{center}
\begin{tabular}{|cc|*{4}{r}|}\hline
\multicolumn{2}{|c|}{Inputs}&\multicolumn{4}{c|}{Output}\\ \hline
0&0&0&0&0&1\\
0&1&0&0&1&0\\
1&0&0&1&0&0\\
1&1&1&0&0&0\\
\hline
\end{tabular}\label{tab:bool}
\end{center}
\end{table}
\subsection{Numerical condition}
\begin{figure}[htb]
\begin{center}
\includegraphics[scale=0.56]{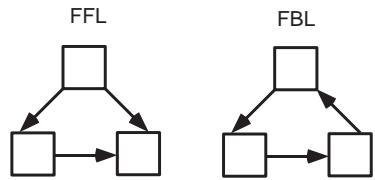}
\caption{Two embedded structures: FFL and FBL stand for feedforward loop and feedback loop, respectively. The loops consist of 3 nodes (squares) and 3 directed edges.}
\label{fig:2subs}
\end{center}
\end{figure}

To investigate the effects of feedback and feedforward loops on the dyamics Boolean networks, we constructed many networks with varying numbers of independent feedback or feedforward loops, where both loops consisted of three nodes and three directed edges [See Fig. \ref{fig:2subs} and Table \ref{tab:con}]. After embedding the specified number of loops, the rest of the directed edges were assigned at random. 
\par
In total, we constructed 9 $\times 10^4$ Boolean networks [See Table \ref{tab:con}] with fixed a network size. We applied 2 $\times 10^3$ initial states to each network. Four different Boolean functions [See Table \ref{tab:bool}] were used in the same frequency [See Table \ref{tab:con}].
\begin{table}[htb]
\begin{center}
\end{center}
\caption{Numerical condition: All networks consist of the same amount of network resources, nodes, directed edges, and the number of Boolean functions. The difference among the generated networks lies in the style of the connections.}
\label{tab:con}
\begin{center}
\begin{tabular}{c|c}
\hline
Size of networks $N$ & 128 nodes\\
Connectivity & For all nodes, $K_{in}$ = $K_{out}$ = 2\\
Boolean function & only AND type [See Table \ref{tab:bool}]\\
Types of loop & FFL, FBL [See Fig. \ref{fig:2subs}]\\
Number of loops & 0, 10, 20, 30, and 40 \\
Number of edges & 256 \\
Number of initial states & 2000 per network\\
Number of realizations & $10^{4}$ in each condition\\
\hline
\end{tabular}
\end{center}
\end{table}
\subsection{Entropy and mutual information}
\begin{figure}[htb]
\begin{center}
\includegraphics[scale=0.5]{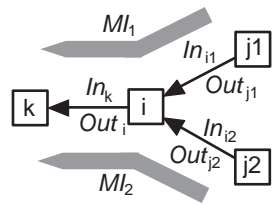}
\caption{Flow of state variables from upstream nodes to downstream. Since the input connectivity for all nodes is 2 (=$K_{in}$), there are two  pathways for mutual information in each node. The input sequence $In_{i1}$ for the node $i$ is the same as the output sequence of an upstream node $Out_{j1}$, and the output sequence of $Out_i$ for the node $i$ is the same as the input sequence of a downstream node $In_k$. When node $i$ has multiple output connections, they have the same binary sequence because state variables in networks are subject to Eq. (\ref{eq:bn}).}
\label{fig:3node-MI}
\end{center}
\end{figure}

\quad We measured the entropy (randomness) and mutual information (coherence) of state variables to characterize the temporal (series) structure of state variables  in the Boolean networks \cite{LANGTON,NDES07,AROB07,PHD02} [See Fig. \ref{fig:3node-MI}].
\subsection{Complexity of state space}
\begin{figure}[htb]
\begin{center}
\includegraphics[scale=0.40]{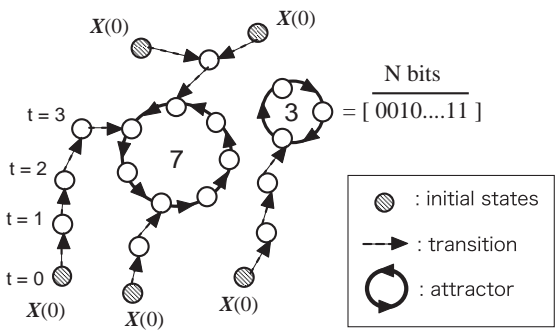}
\caption{Example of a state space: There are different $2^{N}$ states (shown as circles) in the space for each network. The numbers 3 and 7 inside attractors indicate the lengths of the attractors. The numbers also correspond to the size of the basin of attraction.}
\label{fig:state-space}
\end{center}
\end{figure}

To characterize the complexity of state space from initial states [See Fig. \ref{fig:state-space}], we use two measures \cite{ENTROPY}: 
\begin{enumerate}
\item Sum of length of attractors: Each network may contain different number of attractors and the lengths of  the attractors may also vary. The measures define the total length of the attractors in state space.
\item Basin entropy [See Appendix \ref{sec:be} for details]: 
\begin{equation}
H_{Basin}=-\sum_{i} p(i)\log_{2}p(i)
\label{eq:be}
\end{equation}
\end{enumerate}
where, $\sum_{i} p(i)=1$. The two measures indicate the complexity of state space from initial states. According to the definitions, the larger values of two characteristics sgnify higher complexity of the basin of attraction [See Fig. \ref{fig:state-space}].
\subsection{Path length}
To obtain the structual properties of propagating pathway of the state variables, we measured two properties \cite{NDES07,PRE05,MBS07,PHA07}: 
\begin{enumerate}
\item Path length, which is the average number directed edges in the shortest path from a node to all reachable nodes.
\item Average path length, which is the average number of the path lengths for all the nodes.
\end{enumerate}
\section{Results}
\label{sec:nr}
\begin{figure}[hb]
\begin{center}
\includegraphics[scale=0.32]{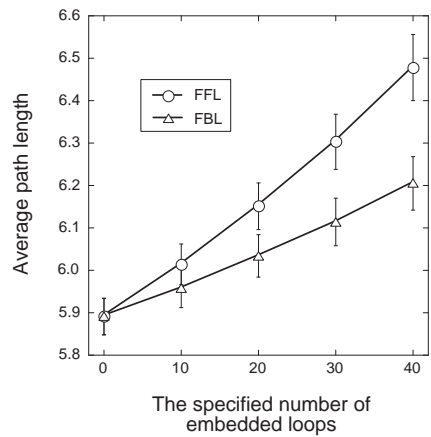}
\caption{Relationship between the number of embedded loops and average path length. Symbols indicate mean $\pm$ SD.}
\label{fig:emb-pl} 
\end{center}
\end{figure}

\begin{figure}[htb]
\begin{center}
\includegraphics[scale=0.32]{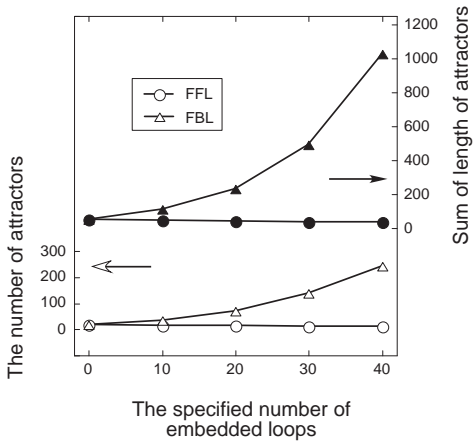}
\caption{Relationship between the number of embedded loops and sum of length of attractors (upper), and the number of attractors (bottom). Symbols indicate mean values.}
\label{fig:emb-atts} 
\end{center}
\end{figure}

\begin{figure}[hb]
\begin{center}
\includegraphics[scale=0.32]{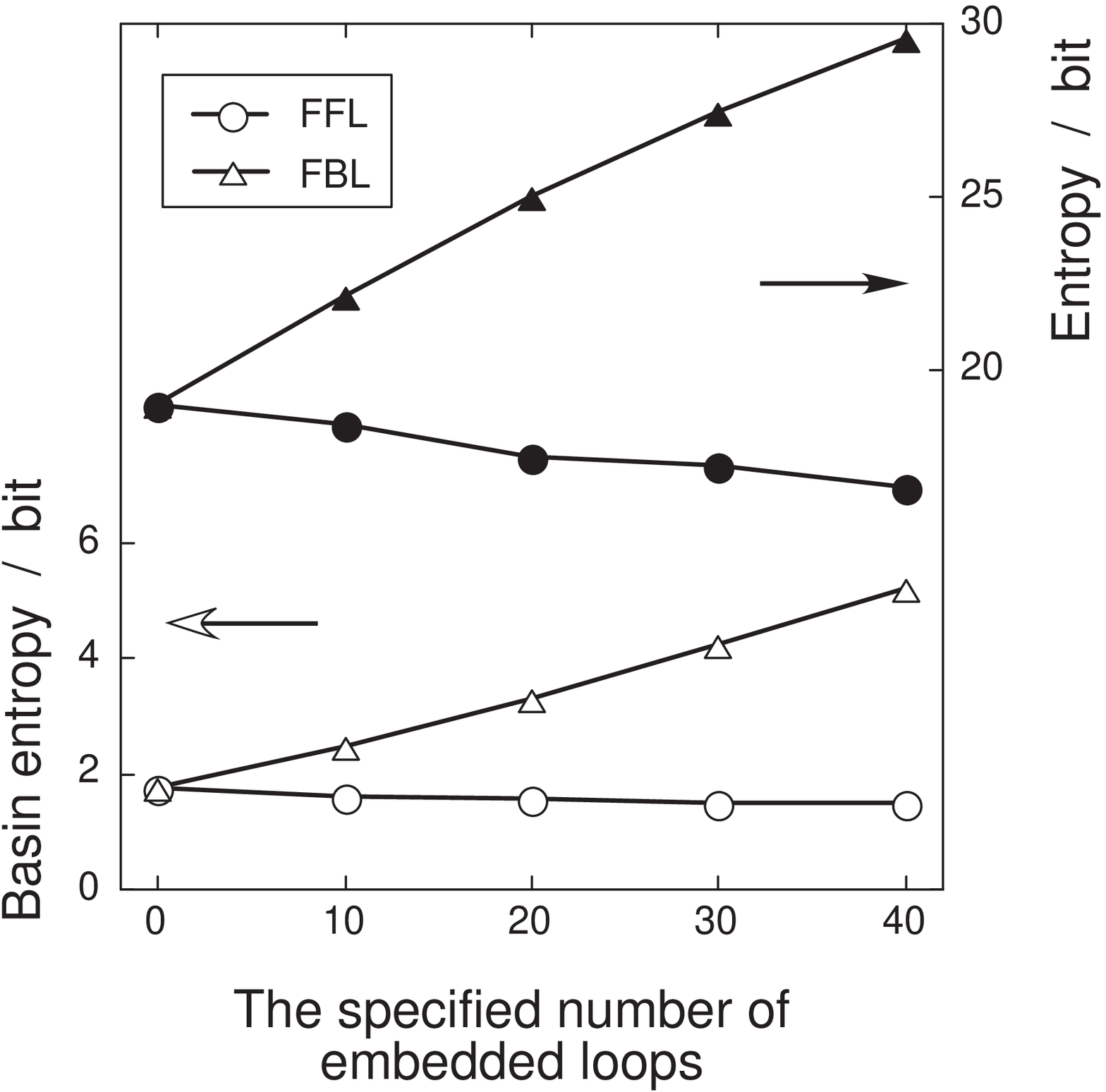}
\caption{Relationship between the number of embedded loops and size of entropy (upper), and basin entropy (bottom). Symbols indicate mean values.}
\label{fig:emb-ens} 
\end{center}
\end{figure}

\begin{figure}[tbh]
\begin{center}
\includegraphics[scale=0.32]{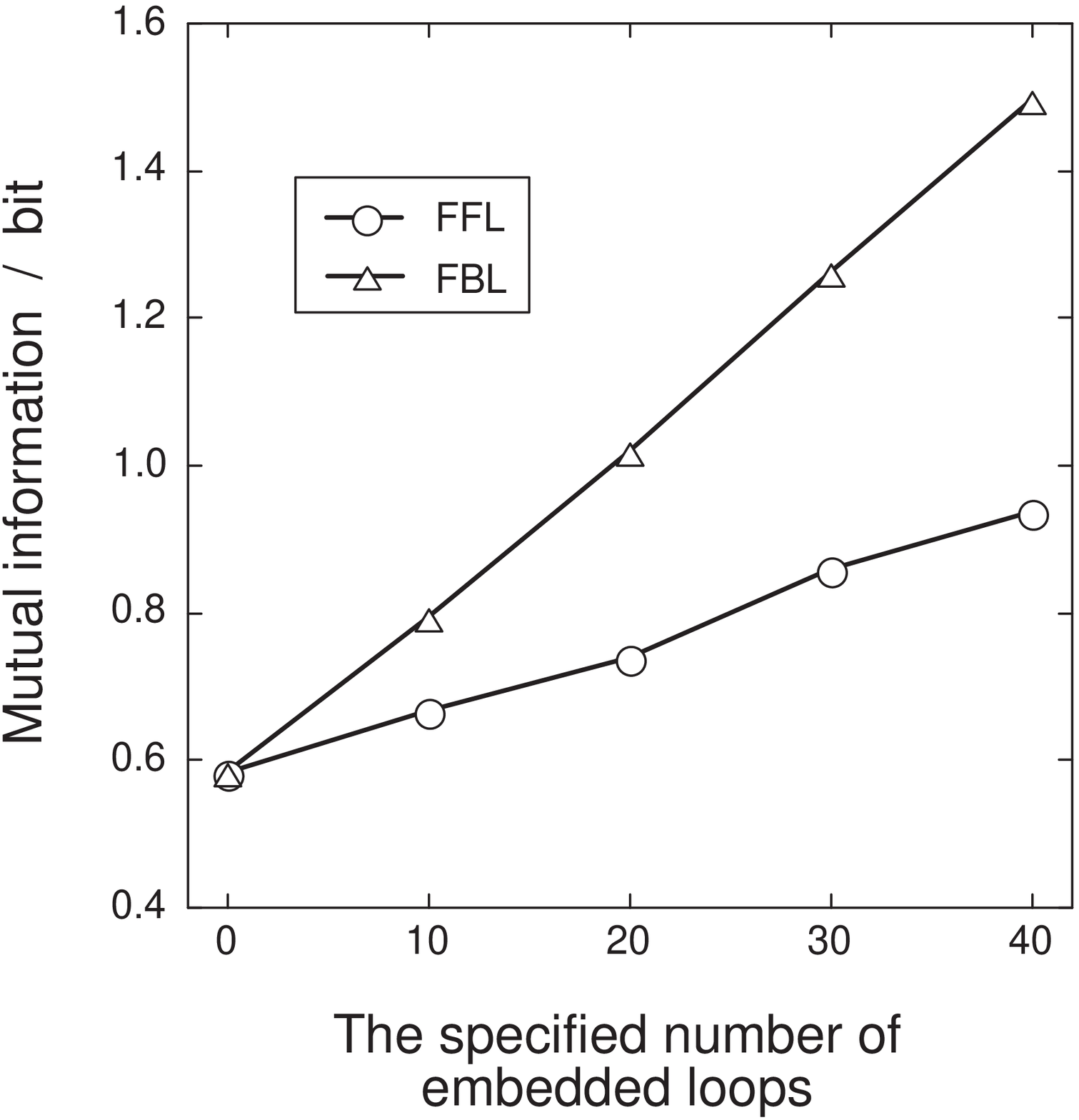}
\caption{Relationship between the number of embedded loops and the size of mutial information. Symbols indicate mean values.}
\label{fig:emb-mi} 
\end{center}
\end{figure}

In total, we obtained 5479157 attractors from 9 $\times 10^{4}$ networks with 1.8 $\times 10^{8}$ intial states. The size of entropy [See Fig. \ref{fig:emb-ens}], mutual information [See Fig. \ref{fig:emb-mi}], total length of attractors [See Fig. \ref{fig:emb-atts}], and basin of entropy [See Fig. \ref{fig:emb-ens}] are measured from the attractors. We also obtained the dependence of average path length on the number of embedded loops [See Fig. \ref{fig:emb-pl}].
\begin{figure}[htb]
\begin{center}
\includegraphics[scale=0.32]{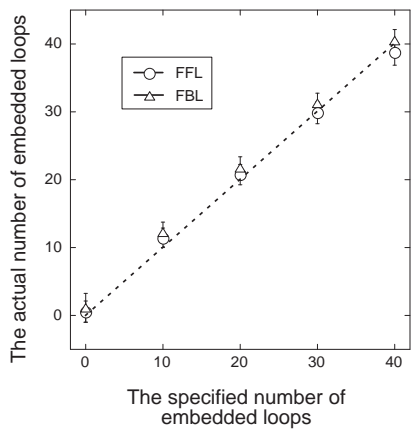}
\caption{Relationship between the specified number of embedded loops and the actual number of embedded loops. Symbols indicate mean $\pm$ SD. Dashed line is given by $y = x$.}
\label{fig:emb-emb}
\end{center}
\end{figure}
\par
Figure \ref{fig:emb-emb} shows that our successful method for embedding loops in Boolean networks. The average path lengths increases with increasing number of embedded loops [See Fig. \ref{fig:emb-pl} and see Appendix \ref{sec:apl}].
Figure \ref{fig:emb-atts} shows the complexity of state-space structures. Feedforward and feedback loops have opposite effects. Please note that both the number of attractors and the sum of the lengths of attractors with feedforward loops slightly decrease. Both entropies in Fig. \ref{fig:emb-ens} indicate opposite effects on different loop structures. 
Unlike Figs. \ref{fig:emb-atts} and \ref{fig:emb-ens}, the size of mutual information increases with increasing number of embedding loops [Fig. \ref{fig:emb-mi}].
\section{Summary and discussion}
We examined the effects of embedding loops on dynamical and topological properties. The networks with more feedback as well as feedforward loops exhibit longer path lengths. As for the dynamical properties, the networks with more feedback loops show a larger number of attractors and greater size of entropy and mutual information, demonstrating that feedback loops increase the complexity of the state space of the networks. In other words, the feedback loops behave as pattern generators of a temporal program of gene expression or entropy generators, and the resultant mutual information (coherence) is driven by the entropy. On the other hand, the networks with more feedforward loops show a smaller number of attractors and size of entropy but larger mutual information. The effects indicate that feedforward loops play a role of stabilizing the state space as well as organizing temporal patterns with less entropy.
\par
The structure of feedback loops resembles that of a repressilator \cite{AG2}, which is a synthetic genetic regulatory network consisting of three genes connected in a feedback loop. The repressilator shows self-sustained oscillations reminiscent of our results from Boolean networks with feedback loops. Other studies \cite{Klemm,Ishihara,Alon} demonstrate that feedforward loops can exhibit temporal and spatial order with differential equations. Our results are similar, as shown in Fig. \ref{fig:emb-mi}.
\par
Control parameters for the dynamics of Boolean networks are input connectivity, $K_{in}$, the size of network, the bias of Boolean functions, and output connectivity distributions \cite{AAM05,MAPC03,PHD02}. In this report, we change only the connection style with the same amount of network resources [See Table \ref{tab:con}]. Nevertheless, Figs. \ref{fig:emb-atts} -- \ref{fig:emb-mi} demonstrate that the internal connection style may well be regarded as a novel control parameter for the dynamics of Boolean networks.
\par
Our results may provide a blueprint for the design of an artificial regulatory gene network \cite{AG2,AG1}, elucidate the role of loop structures in dynamical systems, and provide some insight into the prediction of relationships between complex network structures, behaviors, and functions. Since the currently available biological network data show resultant structures after evolutionary and/or developmental processes, our constructive approach \cite{NDES07,AROB07,PHD02,PRE05,MBS07,PHA07} is one of the promising ways for disentangling natural large-scale complex networks.
\subsection*{Acknowledgements}
We thank Yuichi Ueda for preliminary works. This work was supported by a Grant-in-Aid for Young Scientists (B) No. 18740237 from MEXT, Japan (C.O.). We also thank the authors in http://arxiv.org/abs/0707.3468 is very closely related work.

\section{Appendix}
\subsection{Basin entropy}
\label{sec:be}
The definition of basin entropy deals with the degree of partitioning of state space by basin of attractions. For example, even though the same number of attractors may be found by starting from different inital states, the portions of the number of initial states that follow into those attractors can vary from one network to the other. Eq. (\ref{eq:be}) gives the minimum and maximum entropy when all 2000 different initial states fall into only one attractor and 2000 different attractors, respectively. By the definition, $p(i)$ in Eq. (\ref{eq:be}) is giving as
\begin{equation}
p(i)=\frac{a_{i}}{2000} \nonumber
\end{equation}
\begin{equation}
\sum_{i} a_{i}=2000 \nonumber
\end{equation}
where $a_{i}$ is the number of initial states that reached the i-th attractor, The number 2000 is obtained the numerical condition [See Table \ref{tab:con}].
\subsection{Average path length}
\label{sec:apl}
\begin{figure}[htb]
\begin{center}
\includegraphics[scale=0.45]{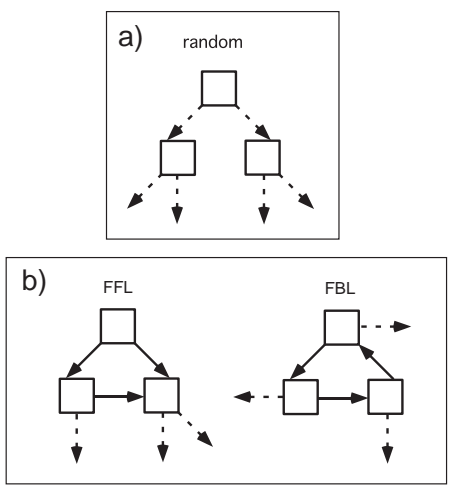}
\caption{Differences in the number of downstream nodes: Dashed arrows indicate output connection from a node a) and from loops b). a) The number of downstream nodes increases almost exponentially with random assignment of edges. b) With embedding loops, the average number of downstream nodes decreases since the number of edges inside the loops becomes large.}
\label{fig:appen-A00}
\end{center}
\end{figure}

Since we generated networks with random assignment of edges between nodes under fixed connectivity, $K_{in}=K_{out}=2$, we can assume a tree-like structure of a nodes in a network (Fig. \ref{fig:appen-A00}a). With this assumption, it can be said that the number of downstream nodes increases exponentially, and the relationships between average path length $L$ and network size $N$ can be expressed as follows  :
\begin{equation}
\frac{2(2^{L}-1)}{2^{1}-1}=N-1 
\label{eq:apl1}
\end{equation}
The average pathlength $L$ can be obtained by transforming Eq. (\ref{eq:apl1}),
\begin{equation}
L=\log_{2}\left(\frac{N-1}{2}+1\right). 
\label{eq:apl2}
\end{equation}
When the number of embedded loops increases, the average actual number of downstream nodes decreases (Fig. \ref{fig:appen-A00}b), leading to the reduction of the common ratio in Eq. (\ref{eq:apl1}). Using the reduced common ratio $x$ where $x$ takes $1 < x < 2$, Eq. (\ref{eq:apl1}) changes to
\begin{equation}
\frac{2(x^{L}-1)}{x^{1}-1}=N-1. 
\label{eq:apl3}
\end{equation}
Similarly, Eq. (\ref{eq:apl3}) can be transformed into
\begin{equation}
L=\log_{x}\left[\frac{(x-1)(N-1)}{2}+1\right]. 
\label{eq:apl4}
\end{equation}

\end{document}